\documentclass[sigconf]{acmart}
\pdfoutput=1

\AtBeginDocument{%
  \providecommand\BibTeX{{%
    \normalfont B\kern-0.5em{\scshape i\kern-0.25em b}\kern-0.8em\TeX}}}

\copyrightyear{2020} 
\acmYear{2020} 
\setcopyright{acmlicensed}
\acmConference[FAT* '20]{Conference on Fairness, Accountability, and Transparency}{January 27--30, 2020}{Barcelona, Spain}
\acmPrice{15.00}
\acmDOI{10.1145/3351095.3372824}
\acmISBN{978-1-4503-6936-7/20/02}

\usepackage{booktabs} 
\usepackage{paralist}
\usepackage[shortlabels]{enumitem}
\usepackage{enumitem}
\usepackage{multirow}
\usepackage{balance}
\usepackage{subfigure}
\usepackage[np, autolanguage]{numprint}
\usepackage{algorithm}
\usepackage{algorithmicx}
\usepackage{xcolor}
\usepackage[noend]{algpseudocode}
\usepackage[skip=1pt]{caption}
\usepackage{hyperref}

\usepackage{acronym}
\acrodef{MC-BRP}{Monte Carlo Bounds for Reasonable Predictions}
\newcommand{\OurCompany}{Ahold Delhaize}
\newcommand{\OurMethod}{\ac{MC-BRP}}
\newcommand{\OurUniversity}{the University of Amsterdam}

\begin{document}

\title[Contrastive Local Explanations for Retail Forecasting]{Why Does My Model Fail? Contrastive Local Explanations for Retail Forecasting}


\author{Ana Lucic}
\orcid{}
\affiliation{%
  \institution{University of Amsterdam}
  \city{Amsterdam}
  \state{Netherlands}
}
\email{a.lucic@uva.nl}

\author{Hinda Haned}
\orcid{}
\affiliation{%
  \institution{Ahold Delhaize}
  \city{Zaandam}
  \state{Netherlands}
}
\email{hinda.haned@aholddelhaize.com}

\author{Maarten de Rijke}
\orcid{0000-0002-1086-0202}
\affiliation{%
  \institution{University of Amsterdam}
  \city{Amsterdam}
  \state{Netherlands}
}
\email{derijke@uva.nl}


\begin{abstract}
In various business settings, there is an interest in using more complex machine learning techniques for sales forecasting. 
It is difficult to convince analysts, along with their superiors, to adopt these techniques since the models are considered to be ``black boxes,'' even if they perform better than current models in use. 
We examine the impact of contrastive explanations about large errors on users' attitudes towards a ``black-box'' model. 
We propose an algorithm, Monte Carlo Bounds for Reasonable Predictions. 
Given a large error, MC-BRP determines (1) feature values that would result in a reasonable prediction, and (2) general trends between each feature and the target, both based on Monte Carlo simulations. 
We evaluate on a real dataset with real users by conducting a user study with 75 participants to determine if explanations generated by MC-BRP help users understand why a prediction results in a large error, and if this promotes trust in an automatically-learned model. 
Our study shows that users are able to answer objective questions about the model's predictions with overall 81.1\% accuracy when provided with these contrastive explanations. 
We show that users who saw MC-BRP explanations understand why the model makes large errors in predictions significantly more than users in the control group. 
We also conduct an in-depth analysis on the difference in attitudes between Practitioners and Researchers, and confirm that our results hold when conditioning on the users' background. 
\end{abstract}



\maketitle

\section{Introduction}
\label{section:1}
As more and more decisions about humans are made by machines, it becomes imperative to understand how these outputs are produced and what drives a model to a particular prediction \citep{riberio-2016-model}. 
As a result, algorithmic interpretability has gained significant interest and traction in the ML community over the past few years \citep{doshi-2017-towards}.
However, there exists considerable skepticism outside of the ML community due to a perceived lack of transparency behind algorithmic predictions, especially when errors are produced \citep{dietvorst-2015-aa}. 
We aim to evaluate the effect of explaining model outputs, specifically large errors, on users' attitudes towards trusting and deploying complex, automatically learned models. 

Further motivation for interpretable ML is provided by significant societal developments. 
Important examples include the recently enacted European General Data Protection Regulation (GDPR), which specifies that individuals will have the right to ``the logic involved in any automatic personal data processing'' \citep{gdpr}. 
In Canada and the United States, this right to an explanation is an integral part of financial regulations, which is why banks have not been able to use high-performing ``black-box'' models to evaluate the credit-worthiness of their customers. 
Instead, they have been confined to easily interpretable algorithms such as decision trees (for segmenting populations) and logistic regression (for building risk scorecards) \citep{khandani-2010-consumer}. 
At NeurIPS 2017, an Explainable ML Challenge was launched to combat this limitation, indicating the finance industry's interest in exploring algorithmic explanations \citep{fico_2017}. 

We use explanations as a mechanism for supporting innovation and technological development while keeping the human ``in the loop'' by focusing on predictive modeling as a tool that aids individuals with a given task. 
Specifically, our interest lies with interpretability in a scenario where users with varying degrees of ML expertise are confronted with large errors in the outcome of predictive models. 
We focus on explaining large errors because people tend to be more curious about unexpected outcomes rather than ones that confirm their prior beliefs~\citep{hilton-1986-knowledge}. 

However, \citet{dietvorst-2015-aa} showed that when users are confronted with errors in algorithmic predictions, they are less likely to use the model. 
Seeing an algorithm make mistakes significantly decreases confidence in the model, and users are more likely to choose a human forecaster instead, even after seeing the algorithm outperform the human \citep{dietvorst-2015-aa}. 
This indicates that prediction mistakes have a significant impact on users' perception of the model. 
By focusing on explaining mistakes, we hope to give insight into this phenomenon of algorithm aversion while also giving users the types of explanations they are interested in seeing. 

Our work was motivated by the needs of analysts at \OurCompany{}, a large Dutch retailer, working on sales forecasting. 
Current models in production are based on simple autoregressive methods, but there is an interest in exploring more complex techniques. 
However, the added complexity comes at the expense of interpretability, which is problematic for \OurCompany{}, especially when a complex model produces a forecast that is very different from the actual target value. 
This leads us to focus on explaining errors in regression predictions in this work. 
However, it should be noted that our method can be extended to classification predictions by defining ``distances'' between classes or by simply defining all errors as large errors. 

We focus on two aspects of explainability in this scenario: the \emph{generation} of explanations of large errors and the corresponding \emph{effectiveness} of these explanations. 
Prior methods for generating explanations fail at generating explanations for large errors because they produce similar explanations for predictions resulting in large errors and those resulting in reasonable predictions (see Table~\ref{table:lime} in Section~\ref{section:4} for an example). 
We propose a method for explaining large prediction errors, called \acfi{MC-BRP}, that shows users: 
\begin{enumerate}[(i)]
\item The required bounds of the most important features in order to have a prediction resulting in a reasonable prediction.
\item The relationship between each of these features and the target.
\end{enumerate}
It should be noted that in our work, we focus on explaining errors \emph{in hindsight}, that is, we examine large errors once they have occurred and are not predicting them in advance without having access to the ground truth. We are also not using these explanations to improve the model, but rather examine the effectiveness of explaining large errors via \OurMethod{} on users' trust in the model and attitudes towards deploying it, as well as their understanding of the explanations. 
We test on a wide range of users, including both Practitioners and Researchers, and analyze the differences in attitudes between these users.  
We also reflect on the process of conducting a user study by outlining some limitations of our study and make some recommendations for future work. 

We address the following research questions: \\
\textbf{RQ1:} \emph{Are the contrastive explanations generated by \OurMethod{} about large errors in predictions
\begin{inparaenum}[(i)]
\item interpretable, or
\item actionable? 
\end{inparaenum}}
More specifically, 
\begin{enumerate}[(i)]
 \item Can contrastive explanations about large errors give users enough information to simulate the model's output (forward simulation)?
 \item Can such explanations help users understand the model such that they can manipulate an observation's input values in order to change the output (counterfactual simulation)?
\end{enumerate}
\textbf{RQ2:} \emph{How does providing contrastive explanations generated by MC-BRP for large errors impact users' perception of the model?} Specifically, we want to investigate the following:
\begin{enumerate}[(i)]
 \item Does being provided with contrastive explanations generated by MC-BRP impact users' understanding of why the model produces errors?
 \item Does it impact their willingness to deploy the model?
 \item Does it impact their level of trust in the model?
 \item Does it impact their confidence in the model's performance?
\end{enumerate}
Consequently, we make the following contributions:
\begin{itemize}
	\item We contribute a method, \OurMethod{}, for generating contrastive explanations specifically for large errors in regression tasks. 
	\item We evaluate our explanations through a user study with \numprint{75} participants in both objective and subjective terms. 
	\item We conduct an analysis on the differences in attitudes between Practitioners and Researchers.  
\end{itemize}

In Section~\ref{section:2} we discuss related work and identify how our problem relates to the current literature. 
In Section~\ref{section:3} we formally describe the methodology of explanations based on \OurMethod{} and in Section~\ref{section:4} we motivate our choice of dataset and describe the user study setup.  
In Section~\ref{section:5} we detail the results of the user study; we conduct further analyses in Section~\ref{section:6}. 
In Section~\ref{section:7} we conclude and make recommendations for future work.


\section{Related Work}
\label{section:2}
\citet{guidotti-2018-survey} compile a survey of current methods in interpretable machine learning and develop a taxonomy for classifying methods using four criteria: 
\begin{itemize}
\item{\textbf{Problem:}}
 \begin{enumerate}[(i)]
 \item{\emph{Model explanations:} interpret black-box model as a whole (globally)}
 \item{\emph{Outcome explanations:} interpret individual black-box predictions (locally)}
 \item{\emph{Inspection:} interpret model behavior through visual representations (globally or locally)}
 \item{\emph{Transparent design:} model is inherently interpretable (globally or locally)}
 \end{enumerate}
\item{\textbf{Model:} neural networks, tree ensembles, SVMs, model-agnostic}
\item{\textbf{Explanator:} decision trees/rules, feature importances, salient masks, sensitivity analysis, partial dependence plots, prototype selection, neuron activation}
\item{\textbf{Data:} tabular, image or text}
\end{itemize}
Based on this schema, our setting is an \emph{outcome explanation} problem for \emph{tree ensembles}. 
We use \emph{sensitivity analysis}, specifically Monte Carlo simulations, on \emph{tabular} data to generate our explanations. 

Existing work on generating outcome explanations specifically for tree ensembles involves finding counterfactual examples \citep{tolomei_interpretable_2017}, identifying influential training samples \citep{sharchilev-2018-finding}, or identifying important features \citep{lundberg_explainable_2019}. 
Importantly, none of these publications are specifically about 
\begin{inparaenum}[(i)]
	\item explaining errors, or
	\item explaining regressions. 
\end{inparaenum} 
On the contrary, these publications are all based on binary classification tasks and the explanations do not necessarily provide insight into prediction mistakes. 

\citet{tolomei_interpretable_2017} propose a method for generating counterfactual examples by identifying decision paths of interest that would result in a different prediction, then traversing down each of these paths and perturbing the instance $x$ such that it satisfies the path in question. 
If this perturbation, $x'$,  
\begin{inparaenum}[(i)]
	\item satisfies the decision path, and
	\item changes the prediction in the overall ensemble, 
\end{inparaenum}
then it is a candidate transformation of $x$. 
After computing all possible candidate transformations by traversing over all paths of interest (i.e., those leading to a different prediction), the candidate transformation with the smallest distance from $x$ is selected as the counterfactual example. 
The explanation, then, is the difference between $x$ and $x'$. 
Although \citet{tolomei_interpretable_2017}'s method also produces contrastive explanations, our method differs from theirs since we are not aiming to identify one counterfactual example, but rather a range of feature values for which the prediction would be different. 
Another difference is that we do not assume full access to the original model. 

\citet{sharchilev-2018-finding} also generate outcome explanations for tree ensembles. 
Their methodology is based on finding influential training samples in order to automatically improve the model, which differs from our work since their explanations are not of a contrastive nature. 
These influential training samples help us understand why a certain class was predicted for a given instance, but they make no reference to the alternative class(es). 
It should be noted that they include a use case on identifying harmful training examples --- ones that contributed to incorrect predictions --- which can be seen as a way to explain errors. 

\citet{lundberg_explainable_2019} propose a method for determining how much each feature contributes to a prediction and present a ranked list of the most important features as the explanation. 
The approach is based on the computationally intensive Shapley values \citep{lundberg_unified_2017}, for which the authors develop a tree-specific approximation. 
This differs from our method since identifying the most important features is only a preliminary step in our pipeline --- our work extends beyond this by including
\begin{inparaenum}[(1)]
\item feature bounds that result in reasonable predictions, and 
\item the relationship between the features and the target as a tool to help users inspect what goes wrong when the prediction error is large.
\end{inparaenum}

\citet{ribeiro-2016-should} also propose a method for identifying local feature importances and this is the one we use in our pipeline. 
Their method, LIME, is model-agnostic and is based on approximating the original model locally with a linear model. 
We share their objective of evaluating users' attitudes towards a model through local explanations but we further specify our task as explaining instances where there are large errors in predictions. 
Based on preliminary experiments, we find that LIME is insufficient for our task setting for two reasons: 
\begin{enumerate}[(i)]
\item For regression tasks, LIME's approximation of the original model is not exact. This ``added'' error can be quite large given that our target is typically of order $10^6$, and this convolutes our definition of a large error.
\item The features LIME deems most important are similar regardless of whether the prediction results in a large error or not, which does not provide any specific insight into why a large error occurs. These experiments are detailed in Section~\ref{section:4}.
\end{enumerate}

\noindent%
Other work on contrastive explanations includes identifying features that should be present or absent in order to justify a classification  \citep{dhurandhar_explanations_2018, ferrari_grounding_2018} or model-agnostic counterfactuals \citep{wachter_counterfactual_2017, russell_efficient_2019}. 
These all differ from our method since they are not specifically about explaining errors. 
Furthermore, the work by \citet{dhurandhar_explanations_2018} and \citet{ferrari_grounding_2018} is based on the binary presence/absence of input features, whereas our method perturbs inputs instead of removing them altogether. 

Our work can also be viewed as a form of outlier detection. 
However, it differs from the standard literature outlined by \citet{pimentel-2014-review} with respect to the objective: we are not necessarily trying to identify outliers in terms of the training data but rather explain instances in the test set whose errors are so large  that they are considered to be anomalies. 

\citet{miller_ijcai_2017} perform a survey of the papers cited in the ``Related Works'' section of the call for the IJCAI 2017 Explainable AI workshop \citep{ijcai-2017-workshop} and find that the majority do not base their methods on the available research about explanations from other disciplines such as philosophy, psychology or cognitive sciences, or evaluate on real users. 
In contrast, our method is rooted in the corresponding philosophical literature \citep{hilton-1990-conversational, lipton-1990-contrastive, hilton-1986-knowledge} and our evaluation is based on a user study.


\section{Method}
\label{section:3}

The intuition behind \OurMethod{} is based on identifying the unusual properties of a particular observation. 
We make the assumption that large errors occur due to unusual feature values in the test set that were not common in the training set. 

Given an observation that results in a large error, \OurMethod{} generates a set of bounds for each feature that would result in a reasonable prediction as opposed to a large error. 
We also include the trend as part of the explanation in order to help users understand the relationship between each feature and the target, and how the input should be changed in order to change the output. 

As pointed out previously, we consider our task of identifying and explaining large errors somewhat similar to that of an outlier detection problem. 
A standard definition of a statistical outlier is an instance that falls outside of a threshold based on the interquartile range. 
A widely used version of this, called Tukey's fences, is defined as follows \citep{tukey-1977-exploratory}:
\[ 
[Q_1 - 1.5(Q_3 - Q_1), Q_3 + 1.5(Q_3 - Q_1)],
\]
where $Q_1$ and $Q_3$ are the first and third quartiles, respectively. 

\begin{definition}\rm
\label{def:large-error}
Let $x$ be an observation in the test set $X$ and let $t$, $\hat{t}$ be the actual and predicted target values of $x$, respectively. Let $\epsilon$ be the corresponding prediction error for $x$, and let $E$ be the set of all errors of $X$. Then $\epsilon$ is a \emph{large error} iff 
\[
\epsilon > Q_3(E) + 1.5(Q_3(E) - Q_1(E)),
\]
where $Q_1(E), Q_3(E)$ are the first and third quartiles of the set of errors, respectively. 
We denote this threshold as $\epsilon_{large}$.
\end{definition}

\noindent%
We can view $X$ in Definition~\ref{def:large-error} as a disjoint union of two sets:
\begin{enumerate}[(i)]
	\item $R$: the set of observations resulting in reasonable predictions, and
	\item $L$: the set of observations resulting in large errors.
\end{enumerate}
We determine the $n$ most important features based on LIME $\Phi^{(x)} = \{\phi_j^{(x)}\}_{j=1}^{n}$, for all $x \in X$. 
It should be noted there exist alternative methods for determining the most important features for a particular prediction \citep{lundberg_unified_2017}, which would also be appropriate. 

\begin{table*}
\caption{An example of an explanation generated by MC-BRP. Here, each of the input values is outside of the range required for a reasonable prediction, which explains why this particular prediction results in a large error. }
\label{table:example}
\begin{tabular}{cp{5.0cm}p{5.8cm}rc}
\toprule
\bf Input & \bf Definition & \bf Trend & \bf Value & \bf Reasonable range\\
\midrule
A & total\_contract\_hrs & As input increases, sales increase
 & 9628.00 & [4140,6565] \\
B & advertising\_costs &  As input increases, sales increase
 & 18160.67 & \phantom{0}[8290,15322] \\
C & num\_transactions & As input increases, sales increase
 & 97332.00 & [51219,75600] \\
D & total\_headcount & As input increases, sales increase
 & 226.00 & \phantom{0}[95,153] \\
E & floor\_surface & As input increases, sales increase
 & 2013.60 & \phantom{0}[972,1725] \\  
\bottomrule
\end{tabular}
\end{table*}
\bigskip

Given $x \in X$, for each $\phi_j^{(x)} \in \Phi^{(x)}$, we determine two sets of characteristics through Monte Carlo simulations:
\begin{enumerate}[(i)]
 \item $[a_{\phi_j^{(x)}}, b_{\phi_j^{(x)}}]$: the bounds for values of $\phi_j^{(x)}$ such that $x \in R$, $x \not\in L$.
 \item $\rho_{\phi_j^{(x)}}$: the relationship between $\phi_j^{(x)}$ and the target we are trying to predict, $t$.
\end{enumerate}
We perturb the feature values for $l \in L$ using Monte Carlo simulations in order to determine what feature values are required to produce a reasonable prediction. 
The algorithm for determining $R'$, the set of Monte Carlo simulations resulting in reasonable predictions, is detailed in Algorithm~\ref{mcbrp}.

In line 3, given $l \in L$, we determine Tukey's fences for each feature in $\Phi^{(l)}$ based on the feature values from $R$. 
This gives us the bounds from which we sample for our feature perturbations. 

In line 5, we randomly sample from these bounds for each $\phi_j^{(l)} \in \Phi^{(l)}$ $m$-times to generate $mn$ versions of our original observation, $l$. 
We call the $i$-th perturbed version $l_i'$, where $i \in \left\{1, \ldots, mn\right\}$. 

In lines 7 and 8, we test the original model $f$ on each $l_i'$, obtain a new prediction, $\hat{t_i'}$, and construct $R'$, the set of perturbations resulting in reasonable predictions. 

Once $R'$ is generated, we compute the mean, standard deviation and Pearson coefficient \citep{swinscow-1997-stats} of the top $n$ features of $l \in L$, $\Phi^{(l)}$, based on this set. 
\bigskip
\begin{algorithm}[h]
    \caption{Monte Carlo simulation: creates a set of perturbed instances resulting in reasonable predictions $R'$ for each large error $l \in L$}
    \label{mcbrp}
    \begin{algorithmic}[1] 
        \Require{instance $l$}
        \Require{set of  $l$'s most important features $\Phi^{(l)}$}
        \Require{`black-box' model $f$}
        \Require{large error threshold $\epsilon_{large}$}
        \Require{number of MC perturbations per feature $m$} 
        \State $R' = \emptyset$ 
            \ForAll{$\phi_j^{(l)}$ in $\Phi^{(l)}$}
                \State $TF(\phi_j^{(l)}$) $\gets$ Tukey's fences for $\phi_j^{(l)}$ \Comment{Based on $R$}
                \For{$i$ in range (0, $m$)}
                    \State $\phi_j'^{(l)} \gets randomsample(TF(\phi_j^{(l)}$))
                    \State $l_i' \gets l_i.replace(\phi_j^{(l)}, \phi_j'^{(l)})$              
                    \State $\hat{t_i'} \gets f(l_i')$ \Comment{New prediction} 
                    \If{$|\hat{t'_i} - t_i| < \epsilon_{large}$}
                        \State {$R' \gets R' \cup l_i'$}
                     \EndIf                    
                \EndFor
            \EndFor
            \Return{$R'$}
    \end{algorithmic}
\end{algorithm}
\begin{definition}\rm
The \emph{trend}, $\rho_{\phi_j^{(x)}}$, of each feature is the Pearson coefficient between each feature $\phi_j^{(x)}$ and the predictions $\hat{t_i'}$ based on the observations in $R'$. It is a measure of linear correlation between two variables \citep{swinscow-1997-stats}. 
\end{definition}

The set of bounds for each feature in $\Phi^{(x)}$ such that $\hat{t}$ results in a reasonable prediction are based on the mean and standard deviation of each $\phi_j^{(x)} \in \Phi^{(x)}$.

\begin{definition}\rm
The \emph{reasonable bounds} for values of each feature $\phi_j$ in $\Phi^{(x)}$, $[a_{\phi_j^{(x)}}, b_{\phi_j^{(x)}}]$, are 
\[
\left[\mu(\phi_j^{(x)})  - \sigma(\phi_j^{(x)}),    \mu(\phi_j^{(x)})  + \sigma(\phi_j^{(x)})\right],
\]
where $\mu(\phi_j^{(x)})$ and $\sigma(\phi_j^{(x)})$ are the mean and standard deviation of each feature, respectively, based on $R'$. 
\end{definition}

We compute the trend and the reasonable bounds for each of the $n$ most important features and present them to the user in a table. 
Table~\ref{table:example} shows an example of an explanation generated by \OurMethod{}; the dataset used for this example is detailed in Section~\ref{section:dataset}.


\section{Experimental Setup} 
\label{section:4}
Current explanation methods mostly serve individuals with ML expertise~\citep{guidotti-2018-survey}, but they should be extended to cater to users outside of the ML community~\citep{miller-2017-explanations}. 
Unlike previous work, our method, \OurMethod{}, generates contrastive explanations by framing the explanation around the prediction error, and aims to help users understand
\begin{inparaenum}[(i)]
\item what contributed to the large error, and 
\item what would need to change in order to produce a reasonable prediction. 
\end{inparaenum}
Presenting explanations in a contrastive manner helps frame the problem and narrows the user's focus regarding the possible outcomes \citep{hilton-1990-conversational,lipton-1990-contrastive}. 

Our explanations are contrastive because they display to the user what would have needed to change in the input order to obtain an alternative outcome from the model --- in other words, why this prediction results in a large error as opposed to a reasonable prediction. 

\subsection{Dataset and model}
\label{section:dataset}
Our task is predicting monthly sales of \OurCompany{}'s stores with \numprint{45} features including financial, workforce and physical store aspects. 
Since not all of our Practitioners have experience with ML, using an internal dataset with familiar features allows them to leverage some of their domain expertise. 
The dataset includes \numprint{45628} observations from \numprint{563} stores, collected at four-week intervals spanning from 2010--2015. 
We split the data by year (training: 2010--2013, test: 2014--2015) to simulate a production environment, and we treat every unique combination of store, interval and year as an independent observation. 
After preprocessing, we have \numprint{21415} and \numprint{12239} observations in our training and test sets, respectively. 
We train the gradient boosting regressor from scikit-learn with the default settings and obtain an $R^{2}$ of \numprint{0.96}.

We verify our assumption that large errors are a result of unusual features values by generating \OurMethod{} explanations for all instances in our test set using $n$ = 5 features and $m$ = \numprint{10000} Monte Carlo simulations. In our dataset, we find that 48\% of instances resulting in large errors have feature values outside the reasonable range for all of the $n$ = 5 most important features, compared to only 24\% of instances resulting in reasonable predictions. 
Although this is not perfect, it is clear that \OurMethod{} produces explanations that are at least somewhat able to distinguish between these two types of predictions. 

\subsection{Why existing solutions are insufficient}
\label{section:lime}
\citet{hilton-2017-social} states that explanations are selective --- it is not necessary or even useful to state all the possible causes that contributed to an outcome. 
The significant part of an explanation is what distinguishes it from the alternative outcome.
If LIME explanations were suitable for our problem, then we would expect to see different features deemed important for instances resulting in large errors compared to those resulting in acceptable errors. 
This would help the user understand why a particular prediction resulted in a large error. 

However, when generating LIME explanations for our test set using $n$ = 5 features, we do not see much of a distinction in the most important features between predictions that result in large errors and those that do not. 
For example, advertising\_costs is one of the top 5 most important features in 18.8\% of instances with large errors and 18.7\% of instances with reasonable predictions. 
These results are summarized in Table~\ref{table:lime}. 
\medskip

\begin{table}[H]
\caption{The top $n=5$ features according to LIME for observations resulting in large errors vs. reasonable predictions.}
\label{table:lime}
\centering
\begin{tabular}{llll}
\toprule
\multicolumn{2}{l}{\textbf{Large errors}} & \multicolumn{2}{l}{\textbf{Reasonable Predictions}} \\
\midrule
advertising\_costs             & 0.188      & advertising\_costs                 & 0.187        \\
total\_contract\_hrs                 & 0.175      & total\_contract\_hrs                    & 0.179        \\
num\_transactions            & 0.151      & num\_transactions               & 0.156        \\
floor\_surface                & 0.124      & total\_headcount                & 0.134        \\
total\_headcount             & 0.123      & floor\_surface                      & 0.122        \\
month                     & 0.109      & month                          & 0.094        \\
mean\_tenure      & 0.046      & mean\_tenure         & 0.046        \\
earnings\_index                & 0.033      & earnings\_index                   & 0.031        \\
\bottomrule
\end{tabular}
\end{table}
\medskip

Furthermore, we originally tried to design our control group user study using explanations from LIME, but found that test users from \OurCompany{} could not make sense of the objective questions about prediction errors because LIME does not provide any insight about errors specifically. 
Given that we could not even ask questions about errors using LIME explanations to users without confusing them, it is clear that LIME is inappropriate for our task. 

\subsection{User study design}
\label{section:studydesign}
We test our method on a real dataset with real users, both from \OurCompany{}. 
We include a short tutorial about predictive modeling along with some questions to check users' understanding as a preliminary component of the study. 
This is because our users are a diverse set of individuals with a wide range of capabilities, including data scientists, human resource strategists, and senior members of the executive team. 
We also include participants from \OurUniversity{} to simulate users who could one day work in this environment. 
In total, we have 75 participants: 44 in the treatment group and 31 in the control group. 

\begin{figure}[t]
\centering
\includegraphics[clip,trim=0mm 0mm 0mm 7.3mm,scale=0.38]{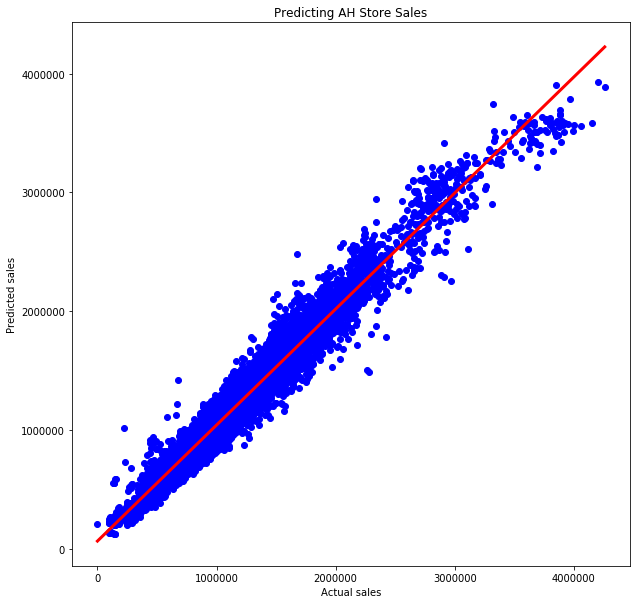}
\caption{The visual description of the model shown to the users: a graph comparing the predicted sales and actual sales based on the original model. The red line depicts perfect predictions.}
\label{fig:pred}
\vspace*{-.5\baselineskip}
\end{figure}

All users are first provided with a visual description of the model: a simple scatter plot comparing the predicted and actual sales (as shown in Figure~\ref{fig:pred}). 
We also show a pie chart depicting the proportion of predictions that result in large errors to give users a sense of how frequently these mistakes occur. 
In our case, this is 4\%. 
Since our users are diverse, we want to make our description of the model as accessible as possible while allowing them to form their own opinions about how well the model performs. 
Participants in the treatment group are shown \OurMethod{} explanations, while those in the control group are not given any explanation. 

\bigskip
The study contains two components, objective and subjective, corresponding to \textbf{RQ1} and \textbf{RQ2}, respectively. 
The objective component is meant to quantitatively evaluate whether or not users understand explanations generated by \OurMethod{}, while the subjective component assesses the effect of seeing the explanation on users' attitudes towards, and perceptions of, the model. 

\smallskip
\begin{table}[h]
\caption{Summary of simulations performed in objective portion of the user study.}
\label{table:simulations}
\centering
\begin{tabular}{lll}
\toprule
\textbf{Type} & \textbf{Provide user with} & \textbf{User's task} \\ 
\midrule
Forward & (1) Input values & Simulate output  \\
 & (2) Explanation & \\ 
\midrule
Counterfactual & (1) Input values & \multirow[t]{3}{2.4cm}{Manipulate input to change output} \\
 & (2) Explanation  &  \\
 & (3) Output  &  \\ 
\bottomrule
\end{tabular}
\end{table}
\smallskip

We base the objective component on \textit{human-grounded metrics}, a framework proposed by \citet{doshi-2017-towards}, where the tasks conducted by users are simplified versions of the original task. 
We modify the original sales prediction task into a binary classification one: we ask users to determine whether or not a prediction will result in a large error, as it seems unreasonable to expect humans to correctly predict retail sales values of order $10^6$. 

\medskip
\begin{table}[h]
\caption{Summary of tasks performed in user study for the treatment and control groups. The subjective questions are asked twice. }
\label{table:study}
\centering
\begin{tabular}{ll}
\toprule
\multicolumn{1}{c}{\textbf{Treatment}} & \multicolumn{1}{c}{\textbf{Control}} \\ 
\midrule
Short modeling tutorial        & Short modeling tutorial         \\
Visual model description            & Visual model description             \\
Subjective questions                & Subjective questions                 \\
Objective questions                 & Dummy questions                      \\
Subjective questions       & Subjective questions   \\
\bottomrule      
\end{tabular}
\end{table}
\medskip

To answer \textbf{RQ1}, we ask users in the treatment group to perform two types of simulations, both suggested by \citet{doshi-2017-towards} and summarized in Table~\ref{table:simulations}. 
The first is \textit{forward simulation}, where we provide participants with the 
\begin{inparaenum}[(i)]
\item input values, and 
\item explanation. 
\end{inparaenum}
We then ask them to simulate the output --- whether or not this prediction will result in a large error. 
The second is \textit{counterfactual simulation}, where we provide participants with the
\begin{inparaenum}[(i)]
\item input values, 
\item explanation, and 
\item output. 
\end{inparaenum}
We then ask them what they would have needed to change in the input in order to change the output. 
In other words, we want participants to determine how the input features can be changed (according to the trend) in order to produce a reasonable prediction as opposed to one that results in large error. 
These objective questions are designed to test whether or not a participant understands the explanations enough to predict or manipulate the model's output. 
We ask every participant in the treatment group to perform two forward simulations and one counterfactual simulation, and we show the same examples to all users. 

For the control group, we found that we could not ask the objective questions in the same way we did for the treatment group. 
This is because the objective component involves simulating the model based on the explanations (see Table~\ref{table:simulations}), which is not possible if the explanations are not provided. 
In fact, we initially left the objective questions in the control group study, but preliminary testing on some users from \OurCompany{} showed that this was confusing and unclear, similar to when we tried using LIME explanations. 
We were concerned this confusion would skew users' perceptions of the model and therefore convolute the results of RQ2. 
Instead, we show participants in the control group the
\noindent
\begin{inparaenum}[(i)]
\item input values, and 
\item output --- whether or not the example resulted in a large error, 
\end{inparaenum}
In this case, we ask them \emph{if they have enough information} to determine why the example does (or does not) result in a large error. 
This serves as a dummy question to engage users with the task without confusing them. 
We cannot ask users in the control group to simulate the model since they do not see the explanations, but we want to mimic the conditions of the treatment group as closely as possible. 
Therefore, \textbf{RQ1}, is solely evaluated on users from the treatment group. 

To answer \textbf{RQ2}, we contrast results from the treatment and control groups. 
We ask both groups of users the same four subjective questions twice, once towards the beginning of the study and once again at the end. 
We ask the questions at the beginning of the study to evaluate the distribution of preliminary attitudes towards the model, based solely on the visual description. 
We ask the questions at the end of the study to evaluate the effectiveness of \OurMethod{} explanations, by comparing the results from the treatment and control groups. 
The questions we devised are based on the user study by \citet{terhoeve-2017-news}. 
Table~\ref{table:study} summarizes the experimental setup for the treatment and control groups. 
Again, the treatment and control groups are treated exactly the same with the exception of the objective questions -- we only ask these to the treatment group since we cannot ask users to simulate the model without giving them the explanation. 

\section{Experimental Results}
\label{section:5}
In this section, we evaluate the explanations generated by \OurMethod{} in terms of 
\begin{inparaenum}[(i)]
	\item objective questions, and 
	\item subjective questions. 
\end{inparaenum}

\subsection{Objective questions}
The results for users' objective comprehension of \OurMethod{} explanations are summarized in Table~\ref{table:objective}. 
We see that explanations generated by MC-BRP are both: 
\begin{inparaenum}[(i)]
\item interpretable and 
\item actionable, 
\end{inparaenum}
with an average accuracy of 81.1\%. 
This answers \textbf{RQ1}. 
When asked to perform forward simulations, the proportion of correct answers was 84.1\% for both questions. 
This indicates that the majority of users were able to interpret the explanations in order to simulate the model's output (\textbf{RQ1:} interpretable). 
When asked to perform counterfactual simulations, the proportion of correct answers was slightly lower at 75.0\%, but still indicates that the majority of users were able to determine how to manipulate the model's input in order to change the output (\textbf{RQ1:} actionable).

\begin{table}[H]
\caption{Results from the objective questions in the user study.}
\label{table:objective}
\centering
\begin{tabular}{ll}
\toprule
\multicolumn{2}{c}{\textbf{Human accuracy}} \\ 
\midrule
Forward simulation 1         & 84.1\%     \\
Forward simulation 2         & 84.1\%     \\
Counterfactual simulation      & 75.0\%     \\ 
\midrule
Average      & 81.1\% \\
\bottomrule
\end{tabular}
\end{table}

\subsection{Subjective questions}
In order to understand the impact of \OurMethod{} explanations on users' attitudes towards the model, we ask them the following subjective questions:
\begin{itemize}
	\item \textbf{SQ1:} I understand why the model makes large errors in predictions.
	\item \textbf{SQ2:} I would support using this model as a forecasting tool.
	\item \textbf{SQ3:} I trust this model.
	\item \textbf{SQ4:} In my opinion, this model produces mostly reasonable outputs.
\end{itemize}

\begin{figure}[H]
 \centering
 \includegraphics[clip,trim=0mm 0mm 0mm 0mm,width=\columnwidth]{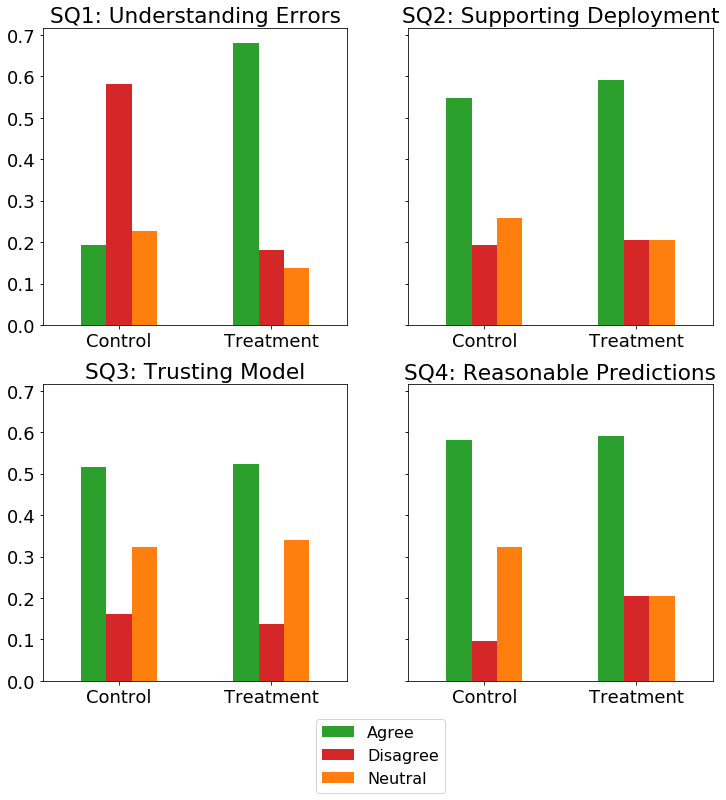}
  \caption{Results from a within-subject study comparing answers between the Treatment (\OurMethod{} explanation) and Control (no explanation) groups.}
 \label{fig:treat-vs-control}
\end{figure}

To ensure our populations did not have different initial attitudes towards the model, we compared their answers on the subjective questions after only showing a visual description of the model. 
The visual description is a graph comparing the predicted sales to the actual sales, which allows users to see the distribution of errors made by the model (see Figure~\ref{fig:pred}). 
We found no statistically significant difference ($\chi^{2}$ test, $\alpha = 0.05$) in initial attitudes towards the model, which allows us to postulate that any difference discovered between the two groups is a result of the treatment they were given (i.e., \OurMethod{} explanation vs. no explanation). 

Figure~\ref{fig:treat-vs-control} shows the distributions of answers to the four subjective questions in the treatment and control groups. 
The difference in distributions is significant for SQ1 ($\chi^{2} = 18.2$, $\alpha = 0.0001$): users in the treatment group agree with the statement more than users in the control group. 
However, we find no statistically significant difference between the two groups for the remaining questions ($\chi^{2}$ test, $\alpha = 0.05$). 
That is, \OurMethod{} explanations help users understand why the model makes large errors in predictions, but do not have an impact on users' trust or confidence in the model, or on their willingness to support its deployment. 
\section{Discussion}
\label{section:6}
Since our original motivation was to provide an explanation system that can be used by analysts at \OurCompany{}, we conducted a more in-depth analysis of the results to determine if there was a difference in attitudes between users depending on their background (e.g., Practitioners from \OurCompany{} or Researchers from \OurUniversity{}). 

\subsection{Comparing attitudes conditioned on background}
\label{section:conditioning}
Table~\ref{table:background} shows the distribution of Practitioners and Researchers in the treatment and control groups. 
Since we have a slight imbalance in background between the treatment and control groups, we test whether or not our results still hold when conditioning on background and confirm that they do. 

Again, we do not find statistically significant differences in initial attitudes towards the model ($\chi^{2}$ test, $\alpha = 0.05$). 
For Researchers, the distribution of answers between treatment and control groups is significantly different for SQ1 ($\chi^{2} = 14.2$, $\alpha = 0.001$), but does not differ for SQ2--SQ4 ($\chi^{2}$ test, $\alpha = 0.05$). 
The same holds for Practitioners: the distributions are significantly different only for SQ1 ($\chi^{2} = 6.94$, $\alpha = 0.05$). 
This is consistent with our results in Section~\ref{section:5}. 
In both cases, users in the treatment group agree with SQ1 more than users in the control group, indicating that \OurMethod{} explanations help users understand why the model makes large errors in predictions, regardless of whether they are Practitioners or Researchers.  
Although the results are statistically significant for both groups, it should be noted that the results hold more strongly for Researchers compared to those for Practitioners, given the $\chi^{2}$ values. 

\begin{table}[H]
\caption{Distribution of Practitioners and Researchers in the treatment and control groups.}
\label{table:background}
\begin{tabular}{lcc}
\toprule
\textbf{Background} & \textbf{Practitioners} & \textbf{Researchers} \\ 
\midrule
Treatment           & 52\%                   & 48\%                 \\
Control             & 58\%                   & 42\%                 \\ 
\bottomrule
\end{tabular}
\end{table}

\subsection{Comparing attitudes in the treatment group}
\begin{figure}[H]
 \centering
 \includegraphics[clip,trim=0mm 0mm 0mm 0mm,width=\columnwidth]{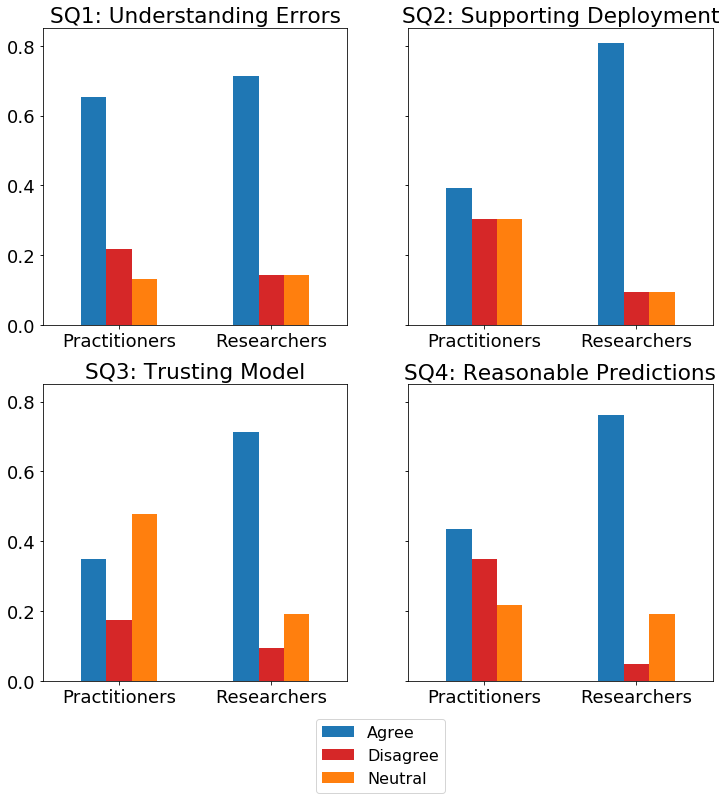}
  \caption{Results from a within-subject study comparing answers between participants who are Practitioners or Researchers (in the treatment group).}
 \label{fig:aca-vs-ind}
\end{figure}

Based on the users who saw the explanations, we compare the distributions of answers between Practitioners and Researchers in Figure~\ref{fig:aca-vs-ind} in order to understand the needs of different types of users. 
We find that there is a significant difference between Practitioners and Researchers for SQ2 ($\chi^{2} = 7.94$, $\alpha = 0.05$), indicating that more Resesearchers are in favor of using the model as a forecasting tool, and less are against it or have a neutral attitude, in comparison to the Practitioners. 
We also find a significant difference for SQ3 ($\chi^{2} = 5.98$, $\alpha = 0.05$): a larger proportion of Researchers trust the model, while the majority of Practitioners have neutral feelings. 
The results for SQ4 are significant as well ($\chi^{2} = 6.86$, $\alpha = 0.05$): 
although the majority of users in both groups believe the model produces reasonable predictions, a larger proportion of the Practitioners disagree with this statement in comparison to the Researchers. 

We see no significant difference between groups for SQ1 ($\chi^{2}$ test, $\alpha = 0.05$), which makes sense given that we showed that \OurMethod{} explanations have a similar effect on both Practitioners and Researchers when comparing users in the treatment and control groups in Section~\ref{section:conditioning}. 

Overall, these results suggest that our user study population is fairly heterogeneous, and that users from different backgrounds have different criteria for deploying or trusting a model, and varying levels of confidence regarding the accuracy of its outcomes. 

\subsection{User study limitations}
Like any user study, ours has some limitations. 
It would have been preferable to distribute users more evenly in terms of the proportion of users in the treatment and control groups, as well as the proportion of Practitioners and Researchers in each of these groups. 
Unfortunately, this was not possible in our case because we recruited participants in two rounds: first for the treatment group, and then afterwards for the control group. 
One option could be to discard some Practitioners in the control group in order to have a better balance in terms of background, but we felt it was more important to have as many users as possible, and it would not be clear how to choose which users to discard. 
Fortunately, we found that our results still hold when conditioning on background as mentioned in Section~\ref{section:conditioning}. 
In future work, we plan to recruit for both groups at the same time to avoid issues like these. 

We also acknowledge that not having a baseline method to compare to is a limitation of our study. 
In our case, the main issue is that there simply does not exist a method that is specifically for explaining errors in regression predictions, which would make asking questions about errors 
\begin{inparaenum}[(i)]
	\item unfair, and 
	\item confusing, as mentioned in Sections~\ref{section:lime} and ~\ref{section:studydesign}. 
\end{inparaenum}
However, now that \OurMethod{}  exists, it can serve as a baseline for future work on erroneous predictions, which is another contribution of this paper. 

\section{Conclusion}
\label{section:7}
We have proposed a method, \acf{MC-BRP}, that provides users with contrastive explanations about predictions resulting in large errors based on:
\begin{inparaenum}[(i)]
\item the set of bounds for which reasonable predictions would be expected for each of the most important features. 
\item the trend between each of these features and the target.
\end{inparaenum}

\noindent%
Given a large error, \OurMethod{} generates a set of perturbed versions of the original instance that result in reasonable predictions. 
This is done by performing Monte Carlo simulations on each of the features deemed most important for the original prediction. 
For each of these features, we determine the bounds needed for a reasonable prediction based on the mean and standard deviation of this new set of reasonable predictions. 
We also determine the relationship between each feature and the target through the Pearson correlation, and present these to the user as the explanation. 

We evaluate \OurMethod{} both objectively (\textbf{RQ1}) and subjectively (\textbf{RQ2}) by conducting a user study with \numprint{75} real users from \OurCompany{} and \OurUniversity{}. 
We answer \textbf{RQ1} by conducting two types of simulations to quantify how 
\begin{inparaenum}[(i)]
\item interpretable, and 
\item actionable 
\end{inparaenum}
our explanations are. 
Through forward simulations, we show that users are able to interpret \OurMethod{} explanations by simulating the model's output with an average accuracy of 84.5\%. 
Through counterfactual simulations, 
we show that \OurMethod{} explanations are actionable with an accuracy of 76.2\%. 

We answer \textbf{RQ2} by conducting a between-subject experiment with subjective questions. 
The treatment group sees \OurMethod{} explanations, while the control group does not see any explanation. 
We find that explanations generated by \OurMethod{} help users understand why models make large errors in predictions (\textbf{SQ1}), but do not have a significant impact on support in deploying the model (\textbf{SQ2}), trust in the model (\textbf{SQ3}), or perceptions of the model's performance (\textbf{SQ4}). 
These results still hold when conditioning on users' background (Practitioners vs. Researchers). 

We also conduct an analysis on the treatment group to compare results between Practitioners and Researchers. 
We find significant differences for \textbf{SQ2, SQ3} and \textbf{SQ4}, but do not find a significant difference in attitudes for \textbf{SQ1}. 

For future work, we intend to explore allowing a predictive model to abstain from prediction when a particular instance has unusual feature values and determine the impact this has on users' trust, deployment support and perception of the model's performance. 
We also plan to compile a more comprehensive set of subjective questions by using multiple questions to evaluate users' impressions on the same topic.

\subsection*{Reproducibility}
To facilitate the reproducibility of the results reported in this work, our code for the experimental implementation of \OurMethod{} is available at
{\fontfamily{qcr}\selectfont
http://github.com/a-lucic/mc-brp}. 

\pagebreak
\bibliographystyle{ACM-Reference-Format}
\bibliography{fat-errors-ana}


\begin{thebibliography}{27}


\ifx \showCODEN    \undefined \def \showCODEN     #1{\unskip}     \fi
\ifx \showDOI      \undefined \def \showDOI       #1{#1}\fi
\ifx \showISBNx    \undefined \def \showISBNx     #1{\unskip}     \fi
\ifx \showISBNxiii \undefined \def \showISBNxiii  #1{\unskip}     \fi
\ifx \showISSN     \undefined \def \showISSN      #1{\unskip}     \fi
\ifx \showLCCN     \undefined \def \showLCCN      #1{\unskip}     \fi
\ifx \shownote     \undefined \def \shownote      #1{#1}          \fi
\ifx \showarticletitle \undefined \def \showarticletitle #1{#1}   \fi
\ifx \showURL      \undefined \def \showURL       {\relax}        \fi
\providecommand\bibfield[2]{#2}
\providecommand\bibinfo[2]{#2}
\providecommand\natexlab[1]{#1}
\providecommand\showeprint[2][]{arXiv:#2}

\bibitem[\protect\citeauthoryear{Dhurandhar, Chen, Luss, Tu, Ting, Shanmugam,
  and Das}{Dhurandhar et~al\mbox{.}}{2018}]%
        {dhurandhar_explanations_2018}
\bibfield{author}{\bibinfo{person}{Amit Dhurandhar}, \bibinfo{person}{Pin-Yu
  Chen}, \bibinfo{person}{Ronny Luss}, \bibinfo{person}{Chun-Chen Tu},
  \bibinfo{person}{Paishun Ting}, \bibinfo{person}{Karthikeyan Shanmugam},
  {and} \bibinfo{person}{Payel Das}.} \bibinfo{year}{2018}\natexlab{}.
\newblock \showarticletitle{Explanations based on the {Missing}: {Towards}
  {Contrastive} {Explanations} with {Pertinent} {Negatives}}.
\newblock In \bibinfo{booktitle}{\emph{Advances in {Neural} {Information}
  {Processing} {Systems} 31}}. \bibinfo{publisher}{Curran Associates, Inc.},
  \bibinfo{pages}{592--603}.
\newblock


\bibitem[\protect\citeauthoryear{Dietvorst, Simmons, and Massey}{Dietvorst
  et~al\mbox{.}}{2015}]%
        {dietvorst-2015-aa}
\bibfield{author}{\bibinfo{person}{Berkeley~J. Dietvorst},
  \bibinfo{person}{Joseph~P. Simmons}, {and} \bibinfo{person}{Cade Massey}.}
  \bibinfo{year}{2015}\natexlab{}.
\newblock \showarticletitle{Algorithm Aversion: People Erroneously Avoid
  Algorithms after Seeing them Err}.
\newblock \bibinfo{journal}{\emph{Journal of Experimental Psychology}}
  \bibinfo{volume}{144} (\bibinfo{year}{2015}), \bibinfo{pages}{114--126}.
\newblock


\bibitem[\protect\citeauthoryear{Doshi-Velez and Kim}{Doshi-Velez and
  Kim}{2018}]%
        {doshi-2017-towards}
\bibfield{author}{\bibinfo{person}{Finale Doshi-Velez} {and}
  \bibinfo{person}{Been Kim}.} \bibinfo{year}{2018}\natexlab{}.
\newblock \showarticletitle{Considerations for Evaluation and Generalization in
  Interpretable Machine Learning}.
\newblock \bibinfo{journal}{\emph{Explainable and Interpretable Models in
  Computer Vision and Machine Learning}} (\bibinfo{year}{2018}).
\newblock


\bibitem[\protect\citeauthoryear{EU}{EU}{2016}]%
        {gdpr}
\bibfield{author}{\bibinfo{person}{EU}.} \bibinfo{year}{2016}\natexlab{}.
\newblock \showarticletitle{Regulation (EU) 2016/679 of the {European
  Parliament} and of the {Council} of 27 {April} 2016 on the protection of
  natural persons with regard to the processing of personal data and on the
  free movement of such data, and repealing {Directive 95/46/EC} ({General Data
  Protection Regulation})}.
\newblock \bibinfo{journal}{\emph{Official Journal of the European Union}}
  \bibinfo{volume}{L119} (\bibinfo{year}{2016}), \bibinfo{pages}{1--88}.
\newblock


\bibitem[\protect\citeauthoryear{FICO}{FICO}{2017}]%
        {fico_2017}
\bibfield{author}{\bibinfo{person}{FICO}.} \bibinfo{year}{2017}\natexlab{}.
\newblock \bibinfo{title}{Explainable Machine Learning Challenge}.
\newblock
\newblock
\urldef\tempurl%
\url{https://community.fico.com/s/explainable-machine-learning-challenge}
\showURL{%
\tempurl}


\bibitem[\protect\citeauthoryear{Guidotti, Monreale, Turini, Pedreschi, and
  Giannotti}{Guidotti et~al\mbox{.}}{2018}]%
        {guidotti-2018-survey}
\bibfield{author}{\bibinfo{person}{Riccardo Guidotti}, \bibinfo{person}{Anna
  Monreale}, \bibinfo{person}{Franco Turini}, \bibinfo{person}{Dino Pedreschi},
  {and} \bibinfo{person}{Fosca Giannotti}.} \bibinfo{year}{2018}\natexlab{}.
\newblock \showarticletitle{A Survey of Methods for Explaining Black Box
  Models}.
\newblock \bibinfo{journal}{\emph{ACM {Computing} Surveys}}
  \bibinfo{volume}{51} (\bibinfo{year}{2018}).
\newblock


\bibitem[\protect\citeauthoryear{Hendricks, Hu, Darrell, and Akata}{Hendricks
  et~al\mbox{.}}{2018}]%
        {ferrari_grounding_2018}
\bibfield{author}{\bibinfo{person}{Lisa~Anne Hendricks},
  \bibinfo{person}{Ronghang Hu}, \bibinfo{person}{Trevor Darrell}, {and}
  \bibinfo{person}{Zeynep Akata}.} \bibinfo{year}{2018}\natexlab{}.
\newblock \showarticletitle{Grounding {Visual} {Explanations}}.
\newblock In \bibinfo{booktitle}{\emph{Computer {Vision} -- {ECCV} 2018}}.
  Vol.~\bibinfo{volume}{11206}. \bibinfo{publisher}{Springer International
  Publishing}, \bibinfo{address}{Cham}, \bibinfo{pages}{269--286}.
\newblock


\bibitem[\protect\citeauthoryear{Hilton}{Hilton}{1990}]%
        {hilton-1990-conversational}
\bibfield{author}{\bibinfo{person}{Dennis~J. Hilton}.}
  \bibinfo{year}{1990}\natexlab{}.
\newblock \showarticletitle{Conversational Processes and Causal Explanation}.
\newblock \bibinfo{journal}{\emph{Psychological Bulletin}}
  \bibinfo{volume}{107} (\bibinfo{year}{1990}), \bibinfo{pages}{65--81}.
\newblock


\bibitem[\protect\citeauthoryear{Hilton}{Hilton}{2017}]%
        {hilton-2017-social}
\bibfield{author}{\bibinfo{person}{Dennis~J. Hilton}.}
  \bibinfo{year}{2017}\natexlab{}.
\newblock \showarticletitle{Social Attribution and Explanation}.
\newblock \bibinfo{journal}{\emph{The Oxford Handbook of Causal Reasoning}}
  (\bibinfo{year}{2017}).
\newblock


\bibitem[\protect\citeauthoryear{Hilton and Slugoski}{Hilton and
  Slugoski}{1986}]%
        {hilton-1986-knowledge}
\bibfield{author}{\bibinfo{person}{Dennis~J. Hilton} {and}
  \bibinfo{person}{Ben~R. Slugoski}.} \bibinfo{year}{1986}\natexlab{}.
\newblock \showarticletitle{Knowledge-based Causal Attribution: The Abnormal
  Conditions Focus Model}.
\newblock \bibinfo{journal}{\emph{Psychological Review}}  \bibinfo{volume}{93}
  (\bibinfo{year}{1986}), \bibinfo{pages}{75--78}.
\newblock


\bibitem[\protect\citeauthoryear{IJCAI}{IJCAI}{2017}]%
        {ijcai-2017-workshop}
\bibfield{author}{\bibinfo{person}{IJCAI}.} \bibinfo{year}{2017}\natexlab{}.
\newblock \bibinfo{title}{Explainable {AI} Workshop}.
\newblock
\newblock
\urldef\tempurl%
\url{http://home.earthlink.net/~dwaha/research/meetings/ijcai17-xai/}
\showURL{%
\tempurl}


\bibitem[\protect\citeauthoryear{Khandani, Kim, and Lo}{Khandani
  et~al\mbox{.}}{2010}]%
        {khandani-2010-consumer}
\bibfield{author}{\bibinfo{person}{Amir~E. Khandani}, \bibinfo{person}{Adlar~J.
  Kim}, {and} \bibinfo{person}{Andrew~W. Lo}.} \bibinfo{year}{2010}\natexlab{}.
\newblock \showarticletitle{Consumer Credit-risk Models via Machine Learning
  Algorithms}.
\newblock \bibinfo{journal}{\emph{Journal of Banking \& Finance}}
  \bibinfo{volume}{34}, \bibinfo{number}{11} (\bibinfo{year}{2010}),
  \bibinfo{pages}{2767--2787}.
\newblock


\bibitem[\protect\citeauthoryear{Lipton}{Lipton}{1990}]%
        {lipton-1990-contrastive}
\bibfield{author}{\bibinfo{person}{Peter Lipton}.}
  \bibinfo{year}{1990}\natexlab{}.
\newblock \showarticletitle{Contrastive Explanation}.
\newblock \bibinfo{journal}{\emph{Royal Institute of Philosophy Supplement}}
  \bibinfo{volume}{27}, \bibinfo{number}{247-266} (\bibinfo{year}{1990}).
\newblock


\bibitem[\protect\citeauthoryear{Lundberg, Erion, Chen, DeGrave, Prutkin, Nair,
  Katz, Himmelfarb, Bansal, and Lee}{Lundberg et~al\mbox{.}}{2019}]%
        {lundberg_explainable_2019}
\bibfield{author}{\bibinfo{person}{Scott~M. Lundberg}, \bibinfo{person}{Gabriel
  Erion}, \bibinfo{person}{Hugh Chen}, \bibinfo{person}{Alex DeGrave},
  \bibinfo{person}{Jordan~M. Prutkin}, \bibinfo{person}{Bala Nair},
  \bibinfo{person}{Ronit Katz}, \bibinfo{person}{Jonathan Himmelfarb},
  \bibinfo{person}{Nisha Bansal}, {and} \bibinfo{person}{Su-In Lee}.}
  \bibinfo{year}{2019}\natexlab{}.
\newblock \showarticletitle{Explainable {AI} for {Trees}: {From} {Local}
  {Explanations} to {Global} {Understanding}}.
\newblock \bibinfo{journal}{\emph{arXiv preprint arXiv:1905.04610}}
  (\bibinfo{year}{2019}).
\newblock


\bibitem[\protect\citeauthoryear{Lundberg and Lee}{Lundberg and Lee}{2017}]%
        {lundberg_unified_2017}
\bibfield{author}{\bibinfo{person}{Scott~M Lundberg} {and}
  \bibinfo{person}{Su-In Lee}.} \bibinfo{year}{2017}\natexlab{}.
\newblock \showarticletitle{A {Unified} {Approach} to {Interpreting} {Model}
  {Predictions}}.
\newblock In \bibinfo{booktitle}{\emph{Advances in {Neural} {Information}
  {Processing} {Systems}}}. \bibinfo{publisher}{Curran Associates, Inc.},
  \bibinfo{pages}{4765--4774}.
\newblock


\bibitem[\protect\citeauthoryear{Miller}{Miller}{2019}]%
        {miller-2017-explanations}
\bibfield{author}{\bibinfo{person}{Tim Miller}.}
  \bibinfo{year}{2019}\natexlab{}.
\newblock \showarticletitle{Explanation in Artificial Intelligence: Insights
  from the Social Sciences}.
\newblock \bibinfo{journal}{\emph{Artificial Intelligence}}
  \bibinfo{volume}{267} (\bibinfo{year}{2019}), \bibinfo{pages}{1--38}.
\newblock


\bibitem[\protect\citeauthoryear{Miller, Howe, and Sonenberg}{Miller
  et~al\mbox{.}}{2017}]%
        {miller_ijcai_2017}
\bibfield{author}{\bibinfo{person}{Tim Miller}, \bibinfo{person}{Piers Howe},
  {and} \bibinfo{person}{Liz Sonenberg}.} \bibinfo{year}{2017}\natexlab{}.
\newblock \showarticletitle{Explainable {AI}: {Beware} of Inmates Running the
  Asylum Or: {How} {I} Learnt to Stop Worrying and Love the Social and
  Behavioural Sciences}.
\newblock \bibinfo{journal}{\emph{IJCAI Workshop on Explainable Artificial
  Intelligence}} (\bibinfo{year}{2017}).
\newblock


\bibitem[\protect\citeauthoryear{Pimentel, Clifton, Clifton, and
  Tarassenko}{Pimentel et~al\mbox{.}}{2014}]%
        {pimentel-2014-review}
\bibfield{author}{\bibinfo{person}{Marco~A.F. Pimentel},
  \bibinfo{person}{David~A. Clifton}, \bibinfo{person}{Lei Clifton}, {and}
  \bibinfo{person}{Lionel Tarassenko}.} \bibinfo{year}{2014}\natexlab{}.
\newblock \showarticletitle{A Review of Novelty Detection}.
\newblock \bibinfo{journal}{\emph{Signal Processing}}  \bibinfo{volume}{99}
  (\bibinfo{year}{2014}), \bibinfo{pages}{215--249}.
\newblock


\bibitem[\protect\citeauthoryear{Ribeiro, Singh, and Guestrin}{Ribeiro
  et~al\mbox{.}}{2016a}]%
        {riberio-2016-model}
\bibfield{author}{\bibinfo{person}{Marco~Tulio Ribeiro},
  \bibinfo{person}{Sameer Singh}, {and} \bibinfo{person}{Carlos Guestrin}.}
  \bibinfo{year}{2016}\natexlab{a}.
\newblock \showarticletitle{Model-Agnostic Interpretability of Machine
  Learning}.
\newblock \bibinfo{journal}{\emph{ICML Workshop on Human Interpretability in
  Machine Learning}} (\bibinfo{year}{2016}).
\newblock


\bibitem[\protect\citeauthoryear{Ribeiro, Singh, and Guestrin}{Ribeiro
  et~al\mbox{.}}{2016b}]%
        {ribeiro-2016-should}
\bibfield{author}{\bibinfo{person}{Marco~Tulio Ribeiro},
  \bibinfo{person}{Sameer Singh}, {and} \bibinfo{person}{Carlos Guestrin}.}
  \bibinfo{year}{2016}\natexlab{b}.
\newblock \showarticletitle{Why Should {I} Trust You?: {Explaining} the
  Predictions of Any Classifier}. In \bibinfo{booktitle}{\emph{Proceedings of
  the 22nd ACM SIGKDD International Conference on Knowledge Discovery and Data
  Mining}}. ACM, \bibinfo{pages}{1135--1144}.
\newblock


\bibitem[\protect\citeauthoryear{Russell}{Russell}{2019}]%
        {russell_efficient_2019}
\bibfield{author}{\bibinfo{person}{Chris Russell}.}
  \bibinfo{year}{2019}\natexlab{}.
\newblock \showarticletitle{Efficient {Search} for {Diverse} {Coherent}
  {Explanations}}.
\newblock \bibinfo{journal}{\emph{arXiv preprint arXiv:1901.04909}}
  (\bibinfo{date}{Jan.} \bibinfo{year}{2019}).
\newblock


\bibitem[\protect\citeauthoryear{Sharchilev, Ustinovsky, Serdyukov, and
  de~Rijke}{Sharchilev et~al\mbox{.}}{2018}]%
        {sharchilev-2018-finding}
\bibfield{author}{\bibinfo{person}{Boris Sharchilev}, \bibinfo{person}{Yury
  Ustinovsky}, \bibinfo{person}{Pavel Serdyukov}, {and}
  \bibinfo{person}{Maarten de Rijke}.} \bibinfo{year}{2018}\natexlab{}.
\newblock \showarticletitle{Finding Influential Training Samples for Gradient
  Boosted Decision Trees}. In \bibinfo{booktitle}{\emph{Proceedings of the 35th
  International Conference on Machine Learning}}.
\newblock


\bibitem[\protect\citeauthoryear{Swinscow}{Swinscow}{1997}]%
        {swinscow-1997-stats}
\bibfield{author}{\bibinfo{person}{Thomas Douglas~Victor Swinscow}.}
  \bibinfo{year}{1997}\natexlab{}.
\newblock \bibinfo{booktitle}{\emph{Statistics at Square One}}.
\newblock \bibinfo{publisher}{BMJ Publishing Group}.
\newblock


\bibitem[\protect\citeauthoryear{ter Hoeve, Heruer, Odijk, Schuth, Spitters,
  and de~Rijke}{ter Hoeve et~al\mbox{.}}{2017}]%
        {terhoeve-2017-news}
\bibfield{author}{\bibinfo{person}{Maartje ter Hoeve}, \bibinfo{person}{Mathieu
  Heruer}, \bibinfo{person}{Daan Odijk}, \bibinfo{person}{Anne Schuth},
  \bibinfo{person}{Martijn Spitters}, {and} \bibinfo{person}{Maarten de
  Rijke}.} \bibinfo{year}{2017}\natexlab{}.
\newblock \showarticletitle{Do News Consumers Want Explanations for
  Personalized News Rankings?}. In \bibinfo{booktitle}{\emph{FATREC Workshop on
  Responsible Recommendation}}.
\newblock


\bibitem[\protect\citeauthoryear{Tolomei, Silvestri, Haines, and
  Lalmas}{Tolomei et~al\mbox{.}}{2017}]%
        {tolomei_interpretable_2017}
\bibfield{author}{\bibinfo{person}{Gabriele Tolomei}, \bibinfo{person}{Fabrizio
  Silvestri}, \bibinfo{person}{Andrew Haines}, {and} \bibinfo{person}{Mounia
  Lalmas}.} \bibinfo{year}{2017}\natexlab{}.
\newblock \showarticletitle{Interpretable {Predictions} of {Tree}-based
  {Ensembles} via {Actionable} {Feature} {Tweaking}}. In
  \bibinfo{booktitle}{\emph{Proceedings of the 23rd ACM SIGKDD International
  Conference on Knowledge Discovery and Data Mining - KDD '17}}.
  \bibinfo{publisher}{ACM}, \bibinfo{pages}{465--474}.
\newblock


\bibitem[\protect\citeauthoryear{Tukey}{Tukey}{1977}]%
        {tukey-1977-exploratory}
\bibfield{author}{\bibinfo{person}{John~W. Tukey}.}
  \bibinfo{year}{1977}\natexlab{}.
\newblock \bibinfo{booktitle}{\emph{Exploratory Data Analysis}}.
\newblock \bibinfo{publisher}{Addison-Wesley}.
\newblock


\bibitem[\protect\citeauthoryear{Wachter, Mittelstadt, and Russell}{Wachter
  et~al\mbox{.}}{2017}]%
        {wachter_counterfactual_2017}
\bibfield{author}{\bibinfo{person}{Sandra Wachter}, \bibinfo{person}{Brent
  Mittelstadt}, {and} \bibinfo{person}{Chris Russell}.}
  \bibinfo{year}{2017}\natexlab{}.
\newblock \showarticletitle{Counterfactual {Explanations} {Without} {Opening}
  the {Black} {Box}: {Automated} {Decisions} and the {GDPR}}.
\newblock \bibinfo{journal}{\emph{SSRN Electronic Journal}}
  (\bibinfo{year}{2017}).
\newblock


\end{thebibliography}
\balance

\end{document}